\documentclass{PoS}
\usepackage{amsmath}
\usepackage{graphicx}
\usepackage{caption}
\usepackage{subcaption}

\usepackage{tikz-feynman}
\tikzfeynmanset{compat=1.0.0}

\newcommand*\dif{\mathop{}\!\mathrm{d}}

\title{Calculating the Two-photon Contribution to $\pi^0 \rightarrow e^+ e^-$ Decay Amplitude}

\ShortTitle{Calculating the Two-photon Contribution to $\pi^0 \rightarrow e^+ e^-$ Decay Amplitude}

\author{Norman H. Christ\\
       Columbia University\\
       E-mail: \email{nhc@phys.columbia.edu}}

\author{Xu Feng\\
       Peking University\\
       E-mail: \email{xu.feng@pku.edu.cn }}

\author{Luchang Jin\\
       University of Connecticut\\
       E-mail: \email{luchang.jin@uconn.edu}}

\author{Cheng Tu\\
       University of Connecticut\\
       E-mail: \email{cheng.tu@uconn.edu}}

\author{\speaker{Yidi Zhao}\thanks{This work is a part of the USQCD QCD+QED project and is partially supported by US DOE grant \#DE-SC0011941.}\\
        Columbia University\\
        E-mail: \email{yz3210@columbia.edu}}

     \abstract{We develop a new method that allows us to deal with two-photon intermediate states in a lattice QCD calculation. We apply this method to perform a first-principles calculation of the $\pi^0 \rightarrow e^+ e^-$ decay amplitude. Both the real and imaginary parts of amplitude are calculated. The imaginary part is compared with the prediction of optical theorem to demonstrate the effectiveness of this method. Our result for the real part of  decay amplitude is $19.68(52)(1.10) \ \text{eV}$, where the first error is statistical and the second is systematic. } 

\FullConference{37th International Symposium on Lattice Field Theory - Lattice2019\\
		16-22 June 2019\\
		Wuhan, China}

\begin{document}

\setlength{\abovedisplayskip}{1pt} 
\setlength{\belowdisplayskip}{1pt}

\section{Introduction}

In the standard model, there is a class of decays involving a combination of QED and QCD processes that are dominated by long-distances contributions. One example that is of interest and can provide important tests of standard model is the rare Kaon decay $K_L\rightarrow \mu^+ \mu^-$. This decay is well established experimentally~\cite{Nakamura_2010} but includes a large long-distance contribution from two-photon intermediate states which migh be computed using lattice QCD.  One difficulty of carrying out this lattice computation is the presence of intermediate two-photon states which can have a lower energy than the initial Kaon state. In this article, we propose a method to tackle this problem and apply this method to perform a lattice computation on a simpler process, the $\pi^0\rightarrow e^+ e^-$ decay.

\begin{figure}
  \centering
\includegraphics[width=0.4\textwidth]{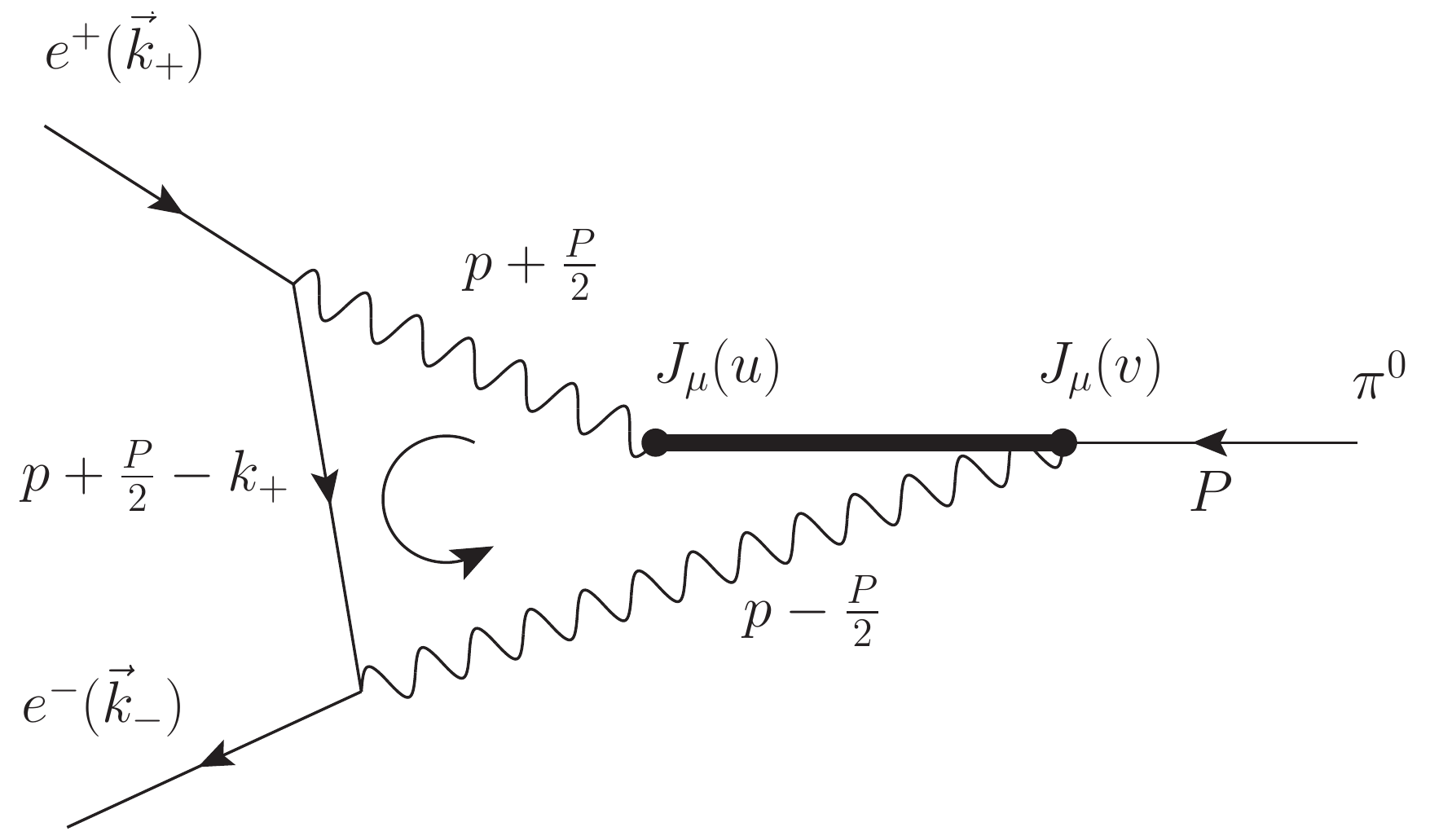}
\caption{The Feynman diagram for $\pi^0 \rightarrow e^+ e^-$ decay.}
\label{fig:pi2ee}
\end{figure}

The $\pi^0 \rightarrow e^+ e^-$ decay process is described by the Feynman diagram in Fig.~\ref{fig:pi2ee}. The decay amplitude can be separated into two parts: the matrix element $\langle 0 | T \bigl\{J_\mu(u) J_\nu(v) \bigr\}| \pi \rangle$ containing contribution from hadronic interaction, and a regular Feynman integral stemming from the internal two-photon and electron loop. The decay amplitude consists of both a real and an imaginary part. The latter is easy to calculate using optical theorem, and gives the well-known unitary bound for the $\pi^0\rightarrow e^+ e^-$ branching ratio. Calculation of the real part of the decay amplitude requires a non-perturbative approach and is the primary goal of this work.

\section{Analytic Continuation Approach}

To calculate $\pi^0 \rightarrow e^+ e^-$ decay on the lattice, we need to solve the problem created by the two-photon intermediate state, whose energy can be lower than initial $\pi^0$ state which implies that a direct Euclidean-space calculation can result in terms which grow exponentially in the time separation. We start with the Minkowski-space expression for the $\pi^0 \rightarrow e^+ e^-$ decay amplitude and write down the matrix element in position space with Minkowski-space time dependence:

\begin{eqnarray}
\mathcal{A} &=&\int \dif^4w \, \langle 0|T\bigl\{J_\mu\bigl(\frac{w}{2}\bigr)J_\nu\bigl(-\frac{w}{2}\bigr)\bigr\}|\pi^0\rangle
\hskip 0.5 in \mbox{\ }  \label{eq:pi-MS} \\
&&\int \dif^4p\ e^{-i p \cdot w} 
                         \left[ \frac{g_{\mu\mu'}}{( p+\frac{P}{2})^2-i\epsilon}\right]
                         \left[ \frac{g_{\nu\nu'}}{( p-\frac{P}{2})^2-i\epsilon}\right] \nonumber \overline{u}(k_-)\gamma_{\mu'}\left[ \frac{\gamma\cdot(p+\frac{P}{2}-k_-) + m_e}{(p+\frac{P}{2}-k_-)^2+m_e^2-i\epsilon}\right]\gamma_{\nu'} v(k_+) \nonumber.
\end{eqnarray}

\begin{figure}[h]
\centering
\includegraphics[width=0.4\textwidth]{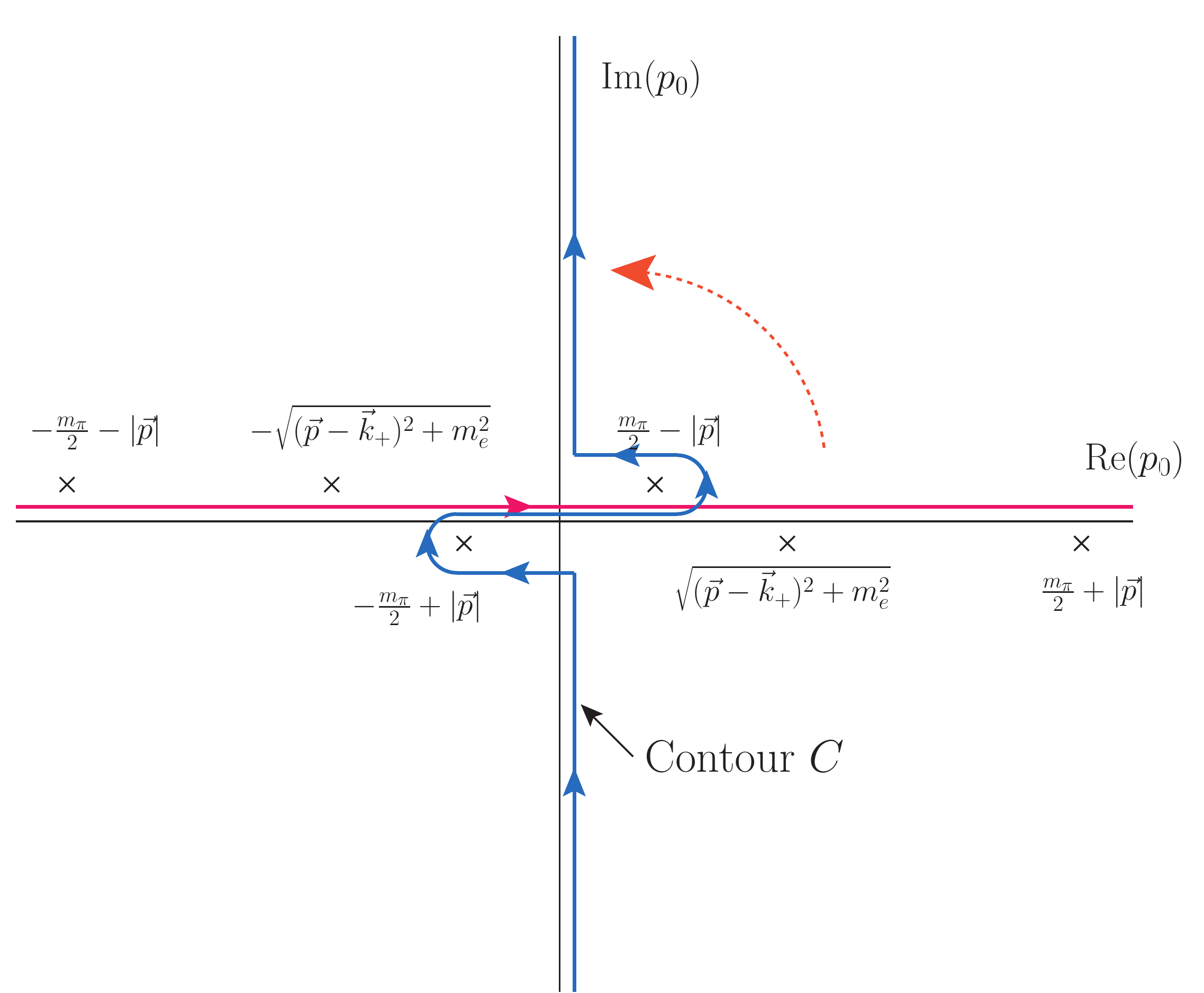}
\caption{A diagram of the complex $p^0$ plane showing integration contours before and after the rotation of the $p^0$ contour.}
\label{fig:p0-contour}
\end{figure}

To convert the matrix element to a Euclidean-space quantity that is calculable on lattice, we rotate the time coordinate $w^0 \rightarrow -i w^0$ and meanwhile perform the opposite rotation on $p^0$, i.e. $p^0 \rightarrow i p^0$. However, because intermediate two-photon state can have less energy than the initial pion state, we have poles in the $p^0$ complex plane that prevent a direct rotation of $p^0$ integration contour from the real axis to the imaginary axis. Thus, as shown in Fig.~\ref{fig:p0-contour}, we deform the contour at the position of the two poles which could cross over imaginary axis when energy of photon is small. It can be shown that the exponential growth of leptonic factor introduced by the deformed contour is overcome by the hadronic matrix matrix which drops faster. More details on this approach can be found in the companion proceeding~\cite{Christ:2019xxx}.

The new integral can be written as:
\begin{eqnarray}
  \mathcal{A} &=& \int \dif^4w \  L_{\mu\nu}(w) H_{\mu\nu}(w) \label{eq:factor} \\
L_{\mu\nu}(w) &=& \int \dif^3p \int_C \dif p_0\  e^{-i \vec p \cdot \vec w}  e^{+p_0 w_0}
       \left[ \frac{\widetilde{g}_{\mu\mu'}}{( p+\frac{P}{2})^2-i\epsilon}\right]
       \left[ \frac{\widetilde{g}_{\nu\nu'}}{( p-\frac{P}{2})^2-i\epsilon}\right] \nonumber \\
&&\hskip 1.0 in \overline{u}(k_-)\gamma_{\mu'}\left[ \frac{\gamma\cdot(p+\frac{P}{2}-k_-) + m_e}{(p+\frac{P}{2}-k_-)^2+m_e^2-i\epsilon}\right]\gamma_{\nu'} v(k_+) \label{eq:L-factor} \\
H_{\mu\nu}(w) &=& 
   \langle 0|T\bigl\{J_\mu\bigl(\frac{w}{2}\bigr)J_\nu\bigl(-\frac{w}{2}\bigr)\bigr\}|\pi^0\rangle_E.
\label{eq:H-factor}
\end{eqnarray}
%
We express the result as the space-time integral of the product of leptonic and hadronic factors. The subscript $E$ on the hadronic matrix element indicates that it is evaluated using Euclidean time dependence and conventions.  The diagonal metric tensor $\widetilde{g}_{\mu\mu'}$ with elements $(1,1,1,i)$ has been introduced to correctly connect the Minkowski conventions for the E\&M currents in the leptonic factor with the Euclidean conventions used in the hadronic matrix element.

\section{Computational Method}




\subsection{Hadronic Factor} \label{sec:hadronic_factor}

Due to Lorentz invariance, when the initial pion state is stationary, the hadronic matrix element can be written as:
\begin{equation}
  H_{\mu\nu}(w) = \epsilon_{0\mu\nu\rho}w^\rho h(w),
  \label{eq:hadron_symm}
\end{equation}
where $h(w)$ is a scalar factor. In our lattice computation, the hadronic factor $H_{\mu\nu}$ can be extracted from the lattice three-point function through the following relationship:
\begin{equation}
  \langle 0 | T \bigl\{J_\mu(x) J_\nu(0) \bigr\}| \pi \rangle =  Z_V^2 \frac{2m_\pi}{N_\pi} \underset{t\rightarrow -\infty}{\lim}  e^{m_\pi |t|} \langle 0 |T \bigl\{J_\mu(x) J_\nu(0) \pi(t) \bigr\}|0 \rangle, 
\end{equation}
where $2m_\pi$ comes from the normalization of pion state, and $N_\pi$ is the normalization factor for pion ground state, i.e. $N_\pi = \langle \pi |\pi(0)|0 \rangle$. The $J_\mu$ and $J_\nu$ operators on the right hand side of the equation are non-conserved local lattice currents and must be multiplied by a renormalization factor $Z_V$.

\begin{figure}
  \centering

  \begin{subfigure}{0.45\textwidth}
    \centering
    \begin{tikzpicture}
      \draw[dashed] (0,-1) -- (0,2);
      \begin{feynman}
        \node [inner sep=0.05cm, circle, fill](u) at (2,1) {};
        \node [inner sep=0.05cm, circle, fill](v) at (2,-0.5) {};
        \node [inner sep=0.05cm, circle, fill](p1) at (0,1.5) {};
        \node [inner sep=0.05cm, circle, fill](p2) at (0,-0.6) {};
        \draw (u) node[right] {$J_\mu(u)$};
        \draw (v) node[right] {$J_\nu(v)$};
        \diagram*{
          {[edges={fermion}]
            (p1) -- (u),
            (u) -- (v),
            (v) -- (p2),
          },
        };
    \end{feynman}
  \end{tikzpicture}
    \caption{Connected diagram}
    \label{fig:connected}
\end{subfigure}
\hfill 
\begin{subfigure}{0.45\textwidth}
  \centering
  \begin{tikzpicture}
    \draw[dashed] (0,-1) -- (0,2);
    \begin{feynman}
      \node [inner sep=0.05cm, circle, fill](u) at (2,1) {};
      \node [inner sep=0.05cm, circle, fill](v) at (2,-0.5) {};
      \node [inner sep=0.05cm, circle, fill](p1) at (0,1.5) {};
      \node [inner sep=0.05cm, circle, fill](p2) at (0,-0.6) {};
      \vertex[above = 1cm of u] (uu);
    \draw (u) node[right] {$J_\mu(u)$};
    \draw (v) node[right] {$J_\nu(v)$};
    \diagram*{
      {[edges={fermion}]
        (p1) -- (v),
        (v) -- (p2),
      },
      {[edges={fermion}]
        (u)  -- [out=135,in=180] (uu) --[scalar,out=0,in=45] (u)
      },
    };
\end{feynman}

\end{tikzpicture}
    \caption{Disconnected diagram}
    \label{fig:disconnected}
\end{subfigure}

\caption{Feynman diagrams for the contractions in the calculation of hadronic factor. The dashed line on the left represents the location of the pion wall source.} \label{fig:feyn}
\end{figure}
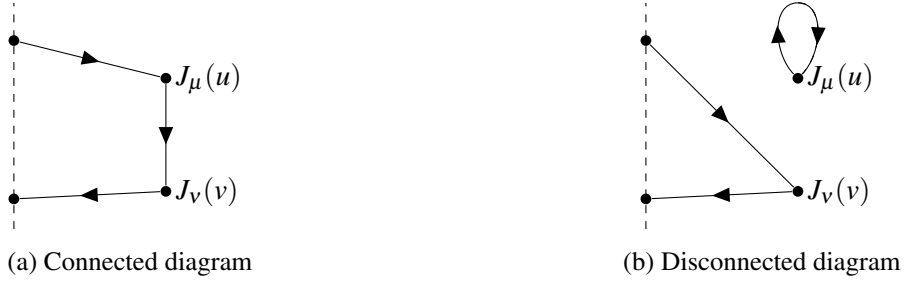

As shown in Fig.~\ref{fig:feyn}, there are two types of diagrams involved in calculating the three point function: connected and disconnected diagrams. The connected diagram is made up of two wall source propagators and one point source propagator. To make sure that the $E\&M$ current is separated far enough from pion operator but not too far such that it crosses the periodic boundary and goes to the other side of pion, we always keep the time difference from the pion wall source at $t$ to the closer current fixed to be a constant $\Delta t$. That means, for every lattice site $x$ in $[-\frac{L}{2}, \frac{L}{2}]$, we always choose a $t$ such that $t = \{ \min(x_0, 0) - \Delta t \} \!\!\!\! \mod L$. The values of $\Delta t$ for each ensembles can be found in Table~\ref{tab:ensembles}.

The disconnected diagram is more difficult to calculate as it involves large noise and would require much more statistics to be calculated accurately. In this work, we make use of the EM loops $\text{Tr}\left[D^{-1}(x, x)\gamma_\mu\right]$ generated with random grid sources from the hadronic vacuum polarization calculation carried out by the RBC/UKQCD  collaboration \cite{Blum:2015you} \cite{Clark:2017wom}. As shown in Section~\ref{sec:results}, this allows us to determine the amplitude of disconnected diagram up to an error of about 60\%.

\subsection{Leptonic Factor}

The leptonic factor $L_{\mu\nu}(w)$ is evaluated by performing the $p^0$ integral using Cauchy's theorem, which leads us to a remaining three-dimensional integral over $\vec{p}$. Note that imaginary part of amplitude is well preserved in this expression. To get the imaginary part, we replace the pole $\frac{1}{|\vec{p}| - M_\pi / 2}$ by a delta function $i \pi \delta(|\vec{p}| - M_\pi / 2)$. The real part of decay amplitude is obtained by taking the principal value.

To further simplify the leptonic factor integral, we make use of the fact that this integral is independent of the direction of the momentum of the outgoing electron $\vec{k}_-$. This enables us to integrate over the angular direction of $\vec{k}_-$ and divide it by $4\pi$. We present the integration result for spatial components here. For the imaginary and real part:
\begin{align}
  L_{ij}^{\text{im}}(w^0, |\vec{w}|) &= \epsilon_{ijk}w^k \frac{1}{|\vec{w}|}\frac{m_e}{M_\pi }\pi \alpha^2 \frac{1}{\beta} \ln \left(\frac{1 + \beta}{1 - \beta} \right) \frac{1}{\frac{M_\pi}{2}|\vec{w}|}\left[\cos(\frac{M_\pi}{2}|\vec{w}|) - \frac{\sin(\frac{M_\pi}{2}|\vec{w}|)}{\frac{M_\pi}{2}|\vec{w}|}\right] \label{eq:L_lep} \\
  L_{ij}^{\text{re}}(w^0, |\vec{w}|) &= 2 \epsilon_{ijk}w^k m_e \frac{\alpha^2}{\pi} \frac{1}{|\vec{w}|} \left(-e^{\frac{M_\pi}{2} |w_0| } \frac{\pi}{M_\pi p_-} \ln \left(\frac{1 + \beta}{1 - \beta}\right) \frac{1}{|\vec{w}|} \right. \\
                                     & \hspace{8em} \int_0^\infty \dif p \frac{e^{-p|w_0|}}{M_\pi - 2p} \left[\cos(p|\vec{w}|) - \frac{\sin(p|\vec{w}|)}{p|\vec{w}|}\right]  \nonumber \\
            & + e^{-\frac{M_\pi}{2} |w_0| } \frac{\pi}{M_\pi p_-} \ln \left(\frac{1 + \beta}{1 - \beta}\right) \frac{1}{|\vec{w}|}\int_0^\infty \dif p \frac{e^{-p|w_0|}}{M_\pi + 2p} \left[\cos(p|\vec{w}|) - \frac{\sin(p|\vec{w}|)}{p|\vec{w}|}\right] \nonumber  \\
            & + \left.  \frac{2\pi}{|\vec{w}|}\int_0^\infty\!\! \dif p \dif \cos\theta \frac{e^{-E_{pe}|w_0|}}{E_{pe}(-M_\pi + 2p_-\cos\theta)(M_\pi + 2p_-\cos\theta)} \left[\cos(p|\vec{w}|) - \frac{\sin(p|\vec{w}|)}{p|\vec{w}|}\right] \right), \nonumber
\end{align}
where $E_{pe}$ is energy of internal electron. The leptonic factor for the real part is now a two-dimensional integral and can be evaluated numerically. Note that only spatial components are shown in this equation, namely, $i, j = (x, y, z)$. The time components do not contribute to this amplitude because of the tensor structure of Eq.~\eqref{eq:hadron_symm}. The numerical integration error is easily controlled to be no more than $0.001\%$. The leptonic factors above are tabulated as functions of $w^0$ and $|\vec{w}|$. Their values on the lattice sites are obtained by linear interpolation.

\section{Results and Analysis} \label{sec:results}

We have performed the lattice computation on four different ensembles, whose parameters are listed in Table~\ref{tab:ensembles}. All ensembles use the Iwasaki gauge action and M\"{o}bius domain wall fermions. For all ensembles except 48I, the dislocation-suppressing-determinant-ratio (DSDR) is also used to reduce the chiral symmetry breaking effects. For each configuration, we have 1024 or 2048 point source propagators whose sources are randomly distributed, and Coulomb gauge-fixed wall source propagators with sources on every time slice. The number of point source propagators for each ensemble is listed in Table~\ref{tab:ensembles}.

\begin{table}[thp]
  \small
  \centering
  \begin{tabular}{|c|c|c|c|c|c|}
    \hline
& 24ID & 32ID & 32IDF & 48I \\
\hline
    $a^{-1}$ (GeV) & 1.015 & 1.015 & 1.37& 1.73 \\
    $m_\pi$ (MeV) & 140 & 140  & 143 & 139 \\
    Configuration separation & 10 & 10  & 10 & 20 \\
    Configurations & 47 & 47  & 61 & 31 \\
    point sources & 1024 & 2048 & 1024 & 1024 \\
    $\Delta t$ & 10 & 10 & 14 & 16 \\
    \hline
  \end{tabular}
  \caption{Table of lattice ensembles used in this work. All ensembles are generated by the RBC/UKQCD collaborations~\protect\cite{Blum:2014tka}. Here, $\Delta t$ is the time difference from the pion wall source at $t$ to the closer current, as explained in Section~\label{sec:hadronic_factor}
  }
  \label{tab:ensembles}
\end{table}

\begin{figure}
  \centering
  \begin{subfigure}{0.43\textwidth}
    \centering
    \includegraphics[width=\textwidth]{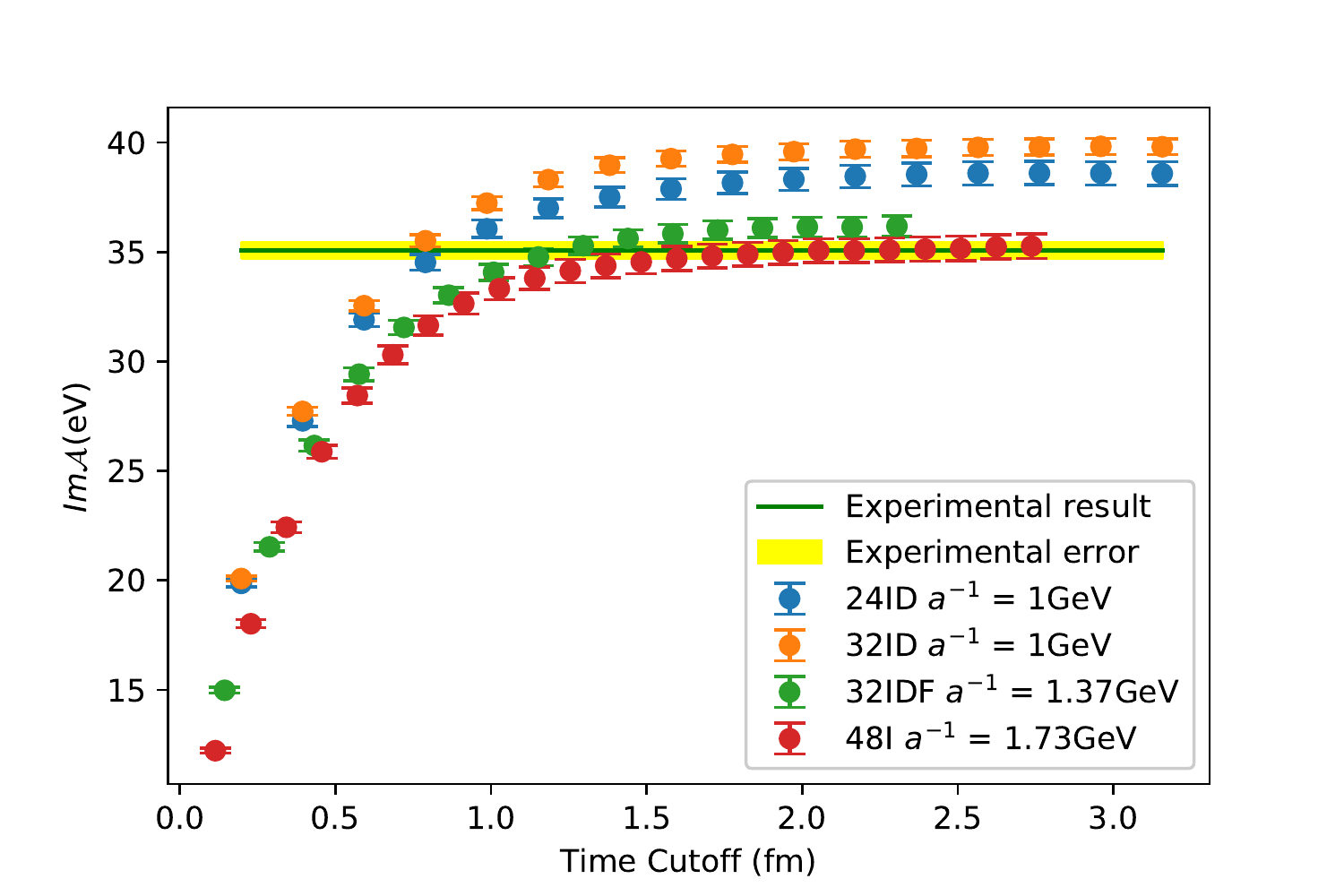}
    \caption{Imaginary Part of Amplitude}
    \label{fig:imag}
  \end{subfigure}%
  \hfill
  \begin{subfigure}{0.43\textwidth}
    \centering
    \includegraphics[width=\textwidth]{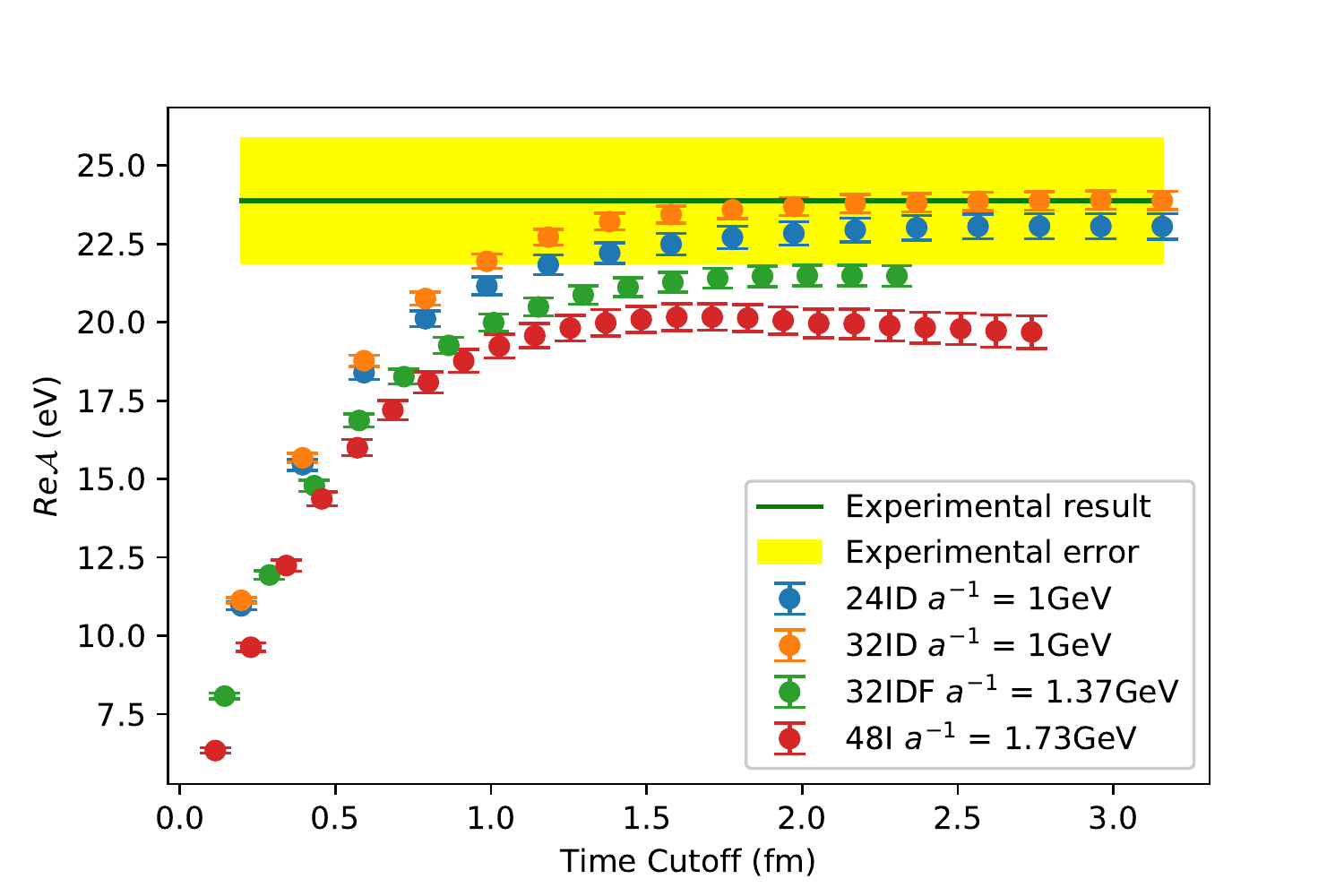}
    \caption{Real Part of Amplitude}
    \label{fig:real}
  \end{subfigure}%

  \caption{Plot of lattice results for decay amplitude v.s. cutoff in time direction.}
  \label{fig:amplitude}
\end{figure}

The results for the real and imaginary parts of the decay amplitude calculated on these four ensembles are listed in Table~\ref{tab:amplitudes}. Note that only the contribution from the connected diagram is included. For the 24ID ensemble, the amplitude from the disconnected diagram is calculated and listed in Table~\ref{tab:24ID_disc}. A plot of amplitude is shown in Fig.~\ref{fig:amplitude}.

\begin{table}[thp]
  \small
  \begin{center}
    \begin{tabular}{|c|c|c|c|c|c|}
      \hline
Source & Im $\mathcal{A}$ (eV) & Re $\mathcal{A}$ (eV) \\ \hline
      24ID &38.58(54) & 23.06(40)  \\
      32ID & 39.80(36) & 23.88(29)  \\
      32IDF & 36.17(47) & 21.48(33)  \\
      48I &35.26(57) & 19.68(52) \\ \hline
      Experiment &35.07(37) & 23.88(1.99) \\
      \hline
    \end{tabular}
  \end{center}
  \caption{Table for comparison among the lattice results and experimental results. The error in parenthesis is statistical. Experimental branching ratio of this decay after radiative corrections is presented in~\cite{PhysRevD.96.014021}. 
  The experimental value for imaginary part is obtained by combining optical theorem and the experimental pion life time. The experimental real part is calculated by subtracting the imaginary part contribution from the total experimental decay rate.}
  \label{tab:amplitudes}
\end{table}

\begin{table}[thp]
  \small
  \begin{center}
    \begin{tabular}{|c|c|c|c|c|c|}
      \hline
Diagram & Im $\mathcal{A}$ (eV) & Re $\mathcal{A}$ (eV) \\ \hline
      24ID &38.58(54) & 23.06(40)  \\
      Disconnected & -1.11(55) & -0.62(40)  \\
      \hline
    \end{tabular}
  \end{center}
  \caption{Contribution to amplitude from connected and disconnected diagrams for the 24ID ensemble. The error in parenthesis is statistical.}
  \label{tab:24ID_disc}
\end{table}

We choose the results obtained from the 48I ensemble because it has the smallest lattice spacing and use our other results to estimate the systematic error presented in Table~\ref{tab:sys-error}. First of all, we did not include the contribution from disconnected diagram. This error can be estimated from the amplitude of disconnected diagram from 24ID ensemble shown in Table~\ref{tab:24ID_disc}. 
The second systematic error is the finite volume error. We have two ensembles with the same lattice spacing but different spatial volume: the 24ID and 32ID ensembles. Assuming that finite volume error behaves as $e^{-m_\pi L}$, we can use the difference of 24ID and 32ID results to estimate the finite volume error for 48ID. 
The third systematic error is finite lattice spacing error. We estimate this error to be 3\% by an order $O((a \Lambda_{\text{QCD}})^2)$ counting with $\Lambda_{\text{QCD}}=300$ MeV.  
Another systematic error arises because the pion masses on the four ensembles are slightly larger than physical $\pi^0$ mass. Using ChPT one can show that the leading order of the hadronic factor is independent of pion mass. We ignore the effects of pion mass on hadronic factor. The imaginary part of the leptonic factor has been worked out analytically in Eq.~\eqref{eq:L_lep}. The major contribution comes from the region where $w$ is small. With Taylor expansion, one can show that the leptonic factor is independent of pion mass in the limit $M_\pi w \ll 1$. Therefore, we ignore the impact of slight pion mass deviation on imaginary part. The real part of the leptonic factor is not analytic. Through numerical experiments, by comparing the difference in the real part of the leptonic part after changing the pion mass, we estimate the error in the leptonic factor in real part to be no more than 1\%.
Additionally, the measurement of renormalization factor in Ref.~\cite{Blum:2014tka}, $Z_V = 0.71076(25)$ also brings an error.
Finally, the errors in leptonic factor numeral integrals are easy to control and are ignored.

We report our final result
\begin{align}
  \text{Im} \mathcal{A} &= 35.26(57)(1.83) \ \text{eV}\\
  \text{Re} \mathcal{A} &= 19.68(52)(1.10) \ \text{eV},
\end{align}
where the first error is statistical and the second is systematic.

\begin{table}[htp]
  \small
  \begin{center}
    \begin{tabular}{|c|c|c|c|c|c|}
      \hline
Sources & Im $\mathcal{A}$ (eV) & Re $\mathcal{A}$ (eV) \\ \hline
Finite volume &  1.09 & 0.74  \\
Finite lattice spacing & 1.06 & 0.59 \\
Disconnected diagram & 1.01 & 0.53 \\
Pion mass &  0 & 0.20  \\
$Z_V$ & 0.025 & 0.014 \\ \hline
Total systematic error& 1.83 & 1.10 \\
      \hline
    \end{tabular}
  \end{center}
  \caption{Sources of systematic error}
  \label{tab:sys-error}
\end{table}

\bibliographystyle{JHEP}
\bibliography{pionee}

%

\end{document}